\documentclass[aps,twocolumn,pra,superscriptaddress,amsmath,amssymb,amsfonts,nofootinbib]{revtex4-1} 
\usepackage[T1]{fontenc}
\usepackage[latin9]{inputenc}
\usepackage{lmodern} 
\usepackage{color}
\usepackage{bbold}
\usepackage{dsfont}
\usepackage{graphicx}
\usepackage[caption=false]{subfig}
\usepackage[colorlinks]{hyperref}
\usepackage[bottom]{footmisc}
\usepackage{enumerate}
\usepackage{float}
\usepackage{mathtools}

\usepackage{footmisc}
\usepackage{empheq}
\usepackage{tikz}
\usepackage{fancyhdr}
\usepackage[bottom]{footmisc}
\usepackage[normalem]{ulem}
\DeclareMathAlphabet\mbc{OMS}{cmsy}{b}{n}
\usepackage{balance}

\begin{document}

\global\long\def\eqn#1{\begin{align}#1\end{align}}
\global\long\def\vec#1{\overrightarrow{#1}}
\global\long\def\ket#1{\left|#1\right\rangle }
\global\long\def\bra#1{\left\langle #1\right|}
\global\long\def\bkt#1{\left(#1\right)}
\global\long\def\sbkt#1{\left[#1\right]}
\global\long\def\cbkt#1{\left\{#1\right\}}
\global\long\def\abs#1{\left\vert#1\right\vert}
\global\long\def\cev#1{\overleftarrow{#1}}
\global\long\def\der#1#2{\frac{{d}#1}{{d}#2}}
\global\long\def\pard#1#2{\frac{{\partial}#1}{{\partial}#2}}
\global\long\def\re{\mathrm{Re}}
\global\long\def\im{\mathrm{Im}}
\global\long\def\dd{\mathrm{d}}
\global\long\def\ddd{\mathcal{D}}

\global\long\def\avg#1{\left\langle #1 \right\rangle}
\global\long\def\mr#1{\mathrm{#1}}
\global\long\def\mb#1{{\mathbf #1}}
\global\long\def\mc#1{\mathcal{#1}}
\global\long\def\tr{\mathrm{Tr}}
\global\long\def\dbar#1{\Bar{\Bar{#1}}}

\global\long\def\nth{$n^{\mathrm{th}}$\,}
\global\long\def\mth{$m^{\mathrm{th}}$\,}
\global\long\def\non{\nonumber}

\newcommand{\orange}[1]{{\color{orange} {#1}}}
\newcommand{\cyan}[1]{{\color{cyan} {#1}}}
\newcommand{\blue}[1]{{\color{blue} {#1}}}
\newcommand{\yellow}[1]{{\color{yellow} {#1}}}
\newcommand{\green}[1]{{\color{green} {#1}}}
\newcommand{\red}[1]{{\color{red} {#1}}}
\global\long\def\todo#1{\orange{{$\bigstar$ \cyan{\bf\sc #1}}$\bigstar$} }

\title{Collective radiation from distant emitters}

\author{Kanupriya Sinha}
\affiliation{Department of Electrical Engineering, Princeton University, Princeton, New Jersey 08544, USA}
\email{kanu@princeton.edu}

\author{Alejandro Gonz\'{a}lez-Tudela}
\affiliation{Institute of Fundamental Physics IFF-CSIC, Calle Serrano 113b, 28006 Madrid, Spain.}
\email{a.gonzalez.tudela@csic.es}

\author{Yong Lu}
\affiliation{Department of Microtechnology and Nanoscience (MC2),
Chalmers University of Technology, SE-412 96 G\"{o}teborg, Sweden}
\email{yongl@chalmers.se}


\author{Pablo Solano}
\affiliation{Departamento de F\'isica, Universidad de Concepci\'on, Casilla 160-C, Concepci\'on, Chile}
\email{psolano@udec.cl}

`\begin{abstract} Waveguides allow for  direct coupling of emitters separated by large distances, offering a path to connect remote quantum systems. However, when facing the distances needed for practical applications, retardation effects due to the finite speed of light are often overlooked. Previous works studied the non-Markovian dynamics of emitters with retardation, but the properties of the radiated field remain mostly unexplored. By considering a toy model of two distant two-level atoms coupled through a waveguide, we observe that the spectrum of the radiated field exhibits non-Markovian features such as linewidth broadening beyond standard superradiance, or narrow Fano resonance-like peaks.  We also show that the  dipole-dipole interaction decays exponentially with distance as a result of retardation, with the range determined by the atomic linewidth. We discuss a proof-of-concept implementation of our results in a  superconducting circuit platform.
\end{abstract}
\maketitle

\section{Introduction}

The interference between coherent radiation processes in an ensemble of atoms leads to collective effects, as first illustrated by Dicke super- and subradiance \cite{Dicke, Haroche}. Collective effects are responsible for a variety of phenomena, relevant in fundamental and applied physics. They can enhance atom-light coupling strengths \cite{Dovzhenko18, Todorov10, Zhang16, Flick18}, which finds applications in quantum information processing  \cite{Hammerer10, Lukin01, Saffman10, Nielsen10}, or can be used to selectively decouple a system from its environment \cite{DFS1, DFS2}, improving the storage and transfer of quantum information \cite{Facchinetti16, Asenjo17, Needham19}. Moreover, collective dipole-dipole interactions, which are responsible for energy exchange between the emitters, can lead to modifications of chemical reactions \cite{Galego16, Hertzog19}, F\"{o}rster energy transfer \cite{Poddubny15, Zhong17, GomezCastano19}, and vacuum-induced energy shifts \cite{Rohlsberger10, Wen19} and forces \cite{Sinha18, Fuchs18}.

Atom-atom interaction strength decreases as the fields propagate away~\cite{Guerin17}, thus collective effects were historically explored in systems with atoms confined to small volumes compared to the radiated wavelengths \cite{Skribanowitz, Gross76, Pavolini85, Devoe96}. However, fields propagating in only one-dimension remove such a constraint, allowing for, in principle, infinite-range interactions \cite{ Solano2017, PabloReview, PabloThesis, Ebongue17, Kato19}. Such one-dimensional systems are therefore an ideal testbed for quantum information applications \cite{ Schoelkopf08, AGT15, Ruostekoski16, Shahmoon, AGT17,Ge18}.  These studies typically employ the Markov approximation \cite{BPbook}. However, when considering collective phenomena over long distances, interference effects are modified as a result of retardation \cite{ Milonni74, superduper1, Milonni76, Berman20}, exhibiting non-Markovian dynamics \cite{Breuer16,deVega17}. Retardation-induced non-Markovianity leads to a variety of phenomena in cavity and half-cavity systems, with dynamics ranging from Rabi oscillations to long-lived non-exponential decay and revivals \cite{Giessen96, Dorner02, Tufarelli13,  Tufarelli14, Carmele13, Cook87, Guimond16}. In collective atom-field interactions, retardation can lead to instantaneous spontaneous emission rates exceeding those of Dicke superradiance \cite{superduper1, Dinc19, Dinc18}, can result in formation of highly delocalized polaritonic modes, referred to as bound states in the continuum (BIC) \cite{superduper1, Hsu16, Fong17, Calajo19, Facchi19, Facchi16, Guo19, Dinc18, SPIE}, and find applications in generating entanglement between distant emitters  \cite{Zheng13,  Pichler16}.

\begin{figure*}[t]
    \centering
    \subfloat[]{\includegraphics[width = 3 in]{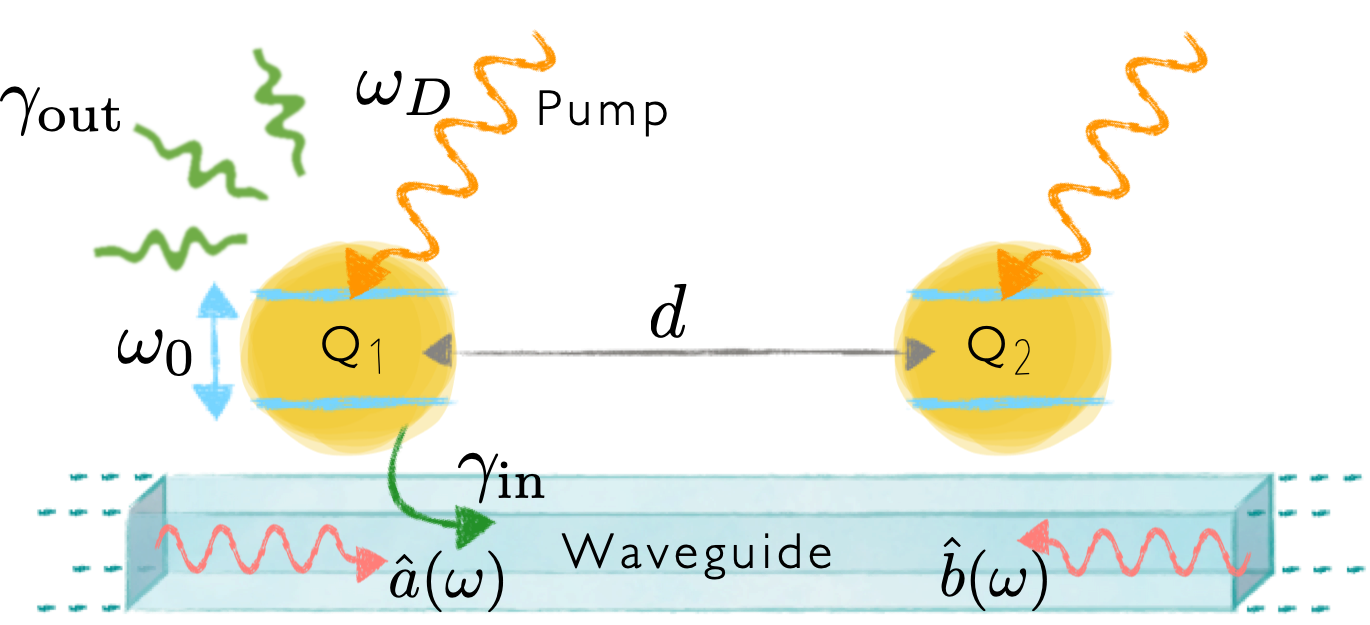}}
    \subfloat[]{\includegraphics[width = 3 in]{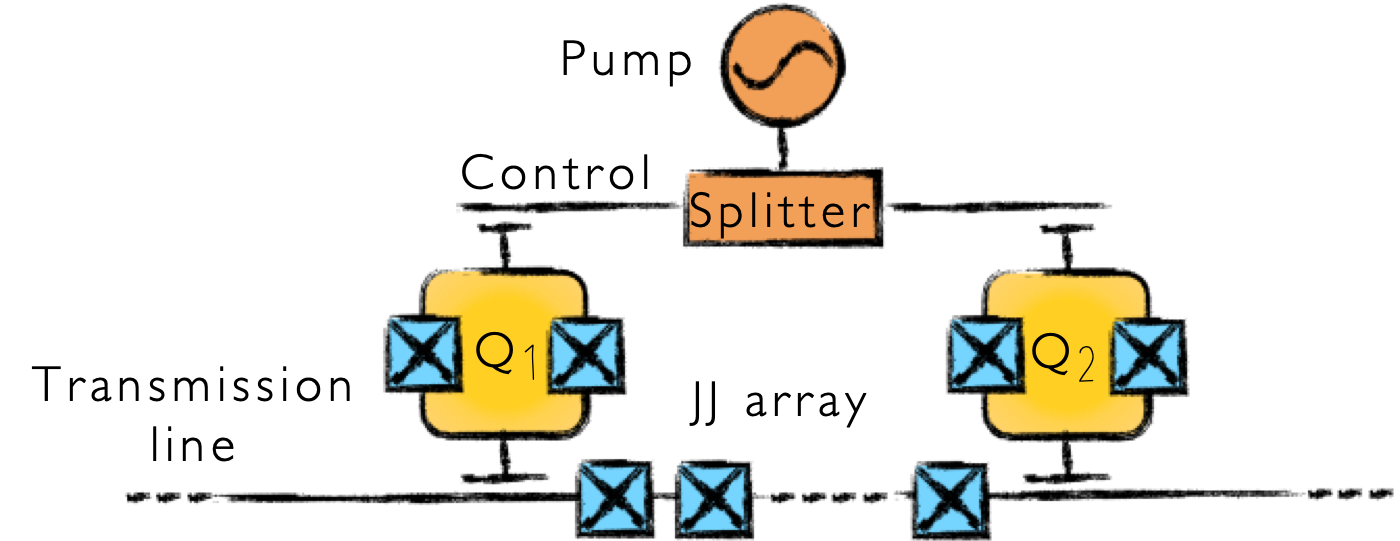}}
    \caption{(a) Schematic representation of two emitters coupled to a one-dimensional waveguide, the qubits have a tunable  resonance frequency of $\omega_0 $ and are separated by a distance $d$ along the waveguide. Each qubit radiates into the waveguide with a rate $\gamma_\mr{in}$, and the dissipation rate outside the waveguide modes is $\gamma_\mr{out}$, with a total emission rate $\gamma \equiv \gamma_\mr{in} + \gamma_{\mr{out}}$. The qubits are pumped by an external driving field simultaneously. (b) Circuit QED implementation of the toy model with two transmon qubits Q1 and Q2 coupled to each other via a Josephson junction array, that is connected to transmission lines on each side \cite{Koch, Yong19, Masluk12}. A split pump field drives the two qubits simultaneously via the control line. Further details of the setup are described in Section\,\ref{superconducting} and the parameter values pertaining to the model are summarized in Table\,\ref{parameters}, which we use throughout the paper to obtain results under realistic conditions.  We assume the pump to be weak and  symmetrically coupled to the two qubits with a Rabi  frequency $\Omega$. }
    \label{schematic}
\end{figure*}

In a recent work~\cite{superduper1}, the authors considered a simple model of two emitters coupled to a one-dimensional waveguide which captures the essential features of collective interactions under retardation. There, the focus was set on the spontaneous emission dynamics of certain emitter states, whereas the properties of the radiated field remained mostly unexplored. In this work, we extend the previous analyses of such a system in several ways: i) considering general initial states and atomic separations; ii) we show that the effective dipole-dipole interaction in the retarded regime decays exponentially with atomic separation; iii) we study the spectrum of the radiated field and unravel the features appearing due to non-Markovian effects; iv) Finally, we study the weakly driven situation to address the preparation of entangled states, considering an implementation of the model in circuit QED platforms, suggesting direct evidences of retarded collective effects that can be experimentally observed.

The paper is organized as follows. In Section\,\ref{model}, we describe the model for the system in consideration as shown in Fig.\,\ref{schematic}(a). In Section \ref{linear} we study the undriven dynamics of the system in the single-excitation subspace. We analyse the response of the system under a weak external drive in Section \ref{drive}, and present a possible superconducting circuit implementation of the model in Section \ref{superconducting} as depicted in Fig.\,\ref{schematic}(b). We summarize our results and  outlook of this work in Section \ref{discussion}.

\section{Model}
\label{model}
Let us consider a system of two distant two-level atoms coupled to a one-dimensional waveguide, as shown in Fig.\,\ref{schematic}\,(a).  One can understand why retardation renders such a system non-Markovian from a simple comparison of time scales. Considering the  subsytems Q1 and Q2 to comprise the system system of interest and the EM field as the bath, the individual relaxation rate of each subsystem into the bath is $\gamma$, with a characteristic relaxation time scale $\tau_{\text{R}} \sim \gamma^{-1}$. The bath mediates the interactions between subsystems at a finite speed $v$, allowing us to define a time scale for bath correlations $\tau_{\text{B}} \sim d/v $. Once the system relaxation rate becomes comparable to the bath correlation time scale, or $\tau_{\text{R}}/\tau_{\text{B}} \sim \gamma d /v \equiv \eta \sim 1$, the Markov approximation is no longer valid \cite{BPbook}\footnote{The  dimensionless parameter $\eta \equiv \gamma d/v$ is also the ratio of the interatomic separation $d$ to the coherence length of a spontaneously emitted photon $(v /\gamma)$. An alternate intuition for  the non-Markovianity in this regime was discussed in \cite{superduper1} in terms of a `superradiance paradox' \cite{Longhi}.}. The parameter $\eta $ captures at least three different sources of non-Markovianity in the context of atom-light interactions \cite{Breuer16, deVega17}: (1) in the strong coupling regime as $\gamma $ increases \cite{Carlos13, Daniele19}, (2) for small propagation velocities \cite{Walther09, Mogilevtsev15} (such as close to a band gap or edge \cite{Thanopulos19, Vats98, John97, AGT2017PRA, AGT2017PRL, Krimer14}), and (3) for large separations $d$, where the interaction between susbsystems is delayed \cite{superduper1,  Whalen17, Grimsmo15, Tufarelli14, Guimond16, Dinc18, Dinc19, SPIE, Carmele20}. Here we explore the system  in a  regime where the interatomic separation is such that $\eta \sim 1$, resulting in retardation-induced non-Markovian dynamics.


The Hamiltonian for a system of two atoms coupled to a waveguide and driven by an external pump field, as depicted in Fig.\,\ref{schematic}\,(a)  is $H = H_{\text{A}}  + H_{\text{F}} + H_{\text{AF}} + H_{\text{AD}}$, where  $H_{\text{A}}  = \sum_{m = 1, 2} \hbar  \omega_0 \hat \sigma_m^+ \hat \sigma_m^ -   $ is the Hamiltonian of the atoms, with $\hat \sigma_m^\dagger$ and $\hat \sigma_m$ as the raising and lowering operators for the $m^{\mr{th}}$ atom. \eqn{ H_{\text{F}} = \int_0 ^\infty \dd\omega \,\hbar \omega \sbkt{\hat{a}^\dagger\bkt{\omega} \hat{a}\bkt{\omega} + \hat{b}^\dagger\bkt{\omega} \hat{b}\bkt{\omega}}   } corresponds to the Hamiltonian for the guided modes of the electromagnetic field, with $\hat a \bkt{\omega}$ and $\hat b \bkt{\omega}$ referring to the right and left propagating modes.  $H_{\text{AF}}$ and $H_{\text{AD}}$ describe the interaction of the atoms with the guided field and an external driving field, respectively.

\vspace{1 cm}

Considering the interaction picture with respect to the free Hamiltonians $H_0 = H_{\text{A}} + H_{\text{F}}$, the  interaction Hamiltonians  $\tilde H_{\text{AF}}\equiv e^{-i H_0 t/\hbar}H_{\text{AF}}e^{i H_0 t/\hbar}$ and  $\tilde H_{\text{AD}}\equiv e^{-i H_0 t/\hbar}H_{\text{AD}}e^{i H_0 t/\hbar}$ can be written as follows.

\begin{widetext}
\eqn{ \label{haf}
\tilde H_{\text{AF}} &= \sum_{m = 1,2} \int_0 ^\infty \dd \omega\,\hbar g\bkt{\omega}\sbkt{\hat{\sigma}_m^\dagger\cbkt{\hat {a}\bkt{\omega}e^{i k x_m } + \hat {b}\bkt{\omega}e^{-i k x_m }}e^{ - i \bkt{\omega - \omega_0 }t} + H.C.}
}
\end{widetext}represents the interaction of the atoms with the waveguide modes within the electric-dipole and rotating wave approximations, wherein $g\bkt{\omega}$ corresponds to the coupling coefficient between the atoms and the field, and $x_m$ is the position of the $m^\mr{th}$ atom \cite{PierreBook}. 
\eqn{\label{had}\tilde H_{\text{AD}} =& \hbar \Omega \sum_{m = 1,2}\sbkt{\hat{\sigma}_m^\dagger e^{-i\bkt{ \omega_{\text{D}}  - \omega_0 }t} + \hat{\sigma}_m e^{i \bkt{\omega_D  - \omega_0 }t}  }
}
is the semi-classical interaction of the emitters with a drive, where $\Omega$ is the Rabi frequency and $\omega_{\text{D}} $ is the drive frequency. One can use the driven Hamiltonian to prepare a particular collective atomic state. We explore this by first solving the problem without the drive for a single excitation subspace in Section \ref{linear} and use the solution to perturbatively calculate the weakly driven dynamics in Section \ref{drive} \cite{Dorner02}.

\section{Dynamics without drive}
\label{linear}
Let us consider the system to be in the single-excitation subspace, where the dynamics is in the linear regime \cite{Calajo16}. The initial state contains one atomic excitation 
\eqn{
\ket{\Psi_0} \equiv\bkt{   \cos\theta \ket{eg}+ \sin \theta e^{i \phi_s} \ket{ge}}\otimes \ket{\cbkt{0}},
}
with the waveguide field in the vacuum state $\ket{\cbkt{0}}$. In the absence of a drive, the total Hamiltonian conserves the number of excitations such that we can use a Wigner-Weisskopf like ansatz to write the state of the atom+field system at a time $t$ as 
\eqn{
\ket{\Psi(t)} = &\left[\sum_{m = 1,2} c_m (t) \hat{\sigma}_m^+ + \int_0 ^\infty \dd\omega \, \left\{c_{a} (\omega,t)  \hat{a}^\dagger\bkt{\omega}( t)\right.\right. \non\\
&\left.\left.+ c_{b} (\omega,t)  \hat{b}^\dagger\bkt{\omega}( t)\right\}\right]\ket{gg}\otimes \ket{\cbkt{0}},
}
where the coefficients $c_m (t) $ refer to the excitation amplitude for the $m^\mr{th}$ emitter and $c_{a(b)} (\omega,t) $ are the excitation amplitudes for the right (left)-propagating mode of the waveguide at frequency $\omega$.\footnote{Notice that the coefficients $c_{a,b} \bkt{\omega, t}$ have dimensions of s$^{-1/2}$ and the total excitation probability for the field modes is $\int  \dd\omega \left(\abs{c_{a} (\omega,t)}^2 + \abs{c_{b} (\omega,t)}^2\right)$.}

The coupled equations of motion due to the atom-field interaction Hamiltonian $\tilde H_{\text{AF}}$ (Eq.\,\eqref{haf}) are
\eqn{\label{ca1}
\dot {c}_a  \bkt{\omega, t}=& -i \sum_{m = 1,2} c_m \bkt{t} g\bkt{\omega} e^{-i \omega x_m /v} e^{i\bkt{ \omega - \omega_0 } t } \\
\label{cb1}
\dot {c}_b  \bkt{\omega, t}=&  -i \sum_{m = 1,2} c_m \bkt{t} g\bkt{\omega} e^{i \omega x_m /v} e^{i\bkt{ \omega - \omega_0 } t }\\
\label{cm1}
\dot {c}_m  \bkt{t}=& -i \int_0 ^\infty \dd\omega \,g\bkt{\omega} e^{-i \bkt{ \omega - \omega_ 0} t} \non\\
&\sbkt{ c_a \bkt{ \omega, t} e^{i \omega x_m/v} +c_b \bkt{ \omega, t} e^{-i \omega x_m/v} }.
}

Formally integrating \eqref{ca1} and \eqref{cb1}, and substituting in \eqref{cm1} gives the equations of motion for the excitation amplitudes of the two atoms:

\eqn{\label{c1dot}
\dot{c}_1\bkt{t} =& -\frac{\gamma}{2} \sbkt{c_1 \bkt{t}  + \beta c_2 \bkt{t - d/v} \Theta \bkt{ t - d/v} e^{i \phi_p} }, \\
\label{c2dot}
\dot{c}_2\bkt{t} =& -\frac{\gamma }{2}\sbkt{c_2 \bkt{t}  + \beta c_1 \bkt{t - d/v} \Theta \bkt{ t - d/v} e^{i \phi_p} },
}
where we have defined $\phi_p \equiv d\omega_0 /v$ as the phase accumulated by the field through its propagation between atoms. Assuming a sufficiently slowly varying density of modes around the atomic resonance, we define the emission into the waveguide as $\gamma_\mr{in} \equiv 4\pi \abs{g\bkt{\omega_0}}^2$. $\gamma = \gamma_\mr{in} +\gamma_\mr{out}  $ is the total spontaneous emission rate of the atom, which includes the radiation outside of the waveguide, which we add phenomenologically. The waveguide coupling efficiency $\beta  = \gamma_\mr{in}/\gamma $ corresponds to the ratio of radiation emitted into the guided modes compared to the total emission. We neglect the effects of field propagation losses, a reasonable approximation for a waveguide based on a JJ array \cite{Masluk12}.

\subsection{Atomic dynamics}
The solutions of the system of coupled delay differential equations given by Eq.\,\eqref{c1dot} and Eq.\,\eqref{c2dot} have the form:
\eqn{\label{c12}
c_ 1\bkt{t} = &K_+ c_+ \bkt{t} +K_- c_- \bkt{t}, \text{ and}\\
c_ 2\bkt{t} = &K_+c_+ \bkt{t}-K_- c_- \bkt{t},
}
where  $K_\pm \equiv \avg{\Psi_\pm\vert\Psi_0}= \bkt{\cos \theta \pm e^{i \phi_s } \sin\theta}/\sqrt2$ is the probability amplitude for the system being initially in the symmetric or anti-symmetric atomic states  $\ket{\Psi_\pm}\equiv\frac{1}{\sqrt 2} \bkt{\ket{eg}\pm \ket{ge}}\otimes \ket{\cbkt{0}} $, and  the functions $c_\pm \bkt{t}$ are the solutions to  the delay differential equation \eqn{\label{dfdt}\der{c_\pm (t) }{t} =-\frac{\gamma }{2}\sbkt{c_\pm(t)\pm \beta e^{i \phi_p }c_\pm \bkt{t - \eta /\gamma}\Theta\bkt{t - \eta/\gamma}}.}
The effect of retardation enters in the second term on the right hand side and is characterized via two parameters, the delay time $\eta/\gamma$ $(= d/v)$ and the waveguide coupling efficiency  times propagation phase factor $\beta e^{i \phi_p }$. The symmetry of the initial state combined with the phase accumulated by the field through propagation determine the overall phase difference of the interference. For example, the initial (anti-)symmetric state $\ket{\Psi_+ } (\ket{\Psi_- })$ with propagation phase $\phi_p  = 2p \pi ~( \phi_p  =(2p+1) \pi) $ is superradiant, while $\ket{\Psi_- } (\ket{\Psi_+ })$ with a propagation phase $\phi_p  = 2p \pi ~(\phi_p  = (2p+1) \pi) $ is subradiant (where $p\in \mathbb{Z}$).
\subsubsection{Solution in terms of Lambert-$W$ functions}

Equation\,\eqref{dfdt} can be solved in terms of Lambert-$W$ functions as \cite{superduper1, Asl03}
    \eqn{\label{fW}
c_\pm (t)  = \frac{1}{\sqrt{2}} \sum_{n= -\infty}^\infty \alpha_n^{(\pm)}e^{ - \gamma_n^{(\pm) }t/2}
}
with
\eqn{\label{gammanpm}
\gamma_n^{(\pm)}&  = {\gamma}\sbkt{1 - \frac{W_n \bkt{ \mp \frac{\eta  }{2}e^{\eta/2}\beta e^{i \phi _ p} }}{\eta /2}}\\ \alpha_n^{(\pm)}& = \sbkt{1  +W_n \bkt{ \mp \frac{\eta  }{2}e^{\eta/2}\beta e^{i \phi _ p} }}^{-1},}
where $W_n (x)$ is the $n^\mr{th}$ branch of the Lambert $W$-function, that often occurs in solutions to delayed-feedback problems \cite{Corless96}. The coefficients $\alpha_n^{\bkt{\pm}}$ and  $\gamma_n^{(\pm)}$ are generally complex valued. 

 We notice that the largest contribution to the sum comes from the terms $n=\cbkt{-1,0,1}$, capturing the qualitative dynamics of the system (see supplemental material in \cite{superduper1}, for example). Particularly, for the symmetric state, $\gamma_{0}^{(+)}$ is real valued for $\eta<\eta_c$, where $\eta _c \equiv 2W_0(\frac{1}{\beta e})$ is defined as a critical distance between the emitters  below which there are no oscillations in the atomic dynamics \cite{superduper1}. Nonetheless, higher order  terms are necessary to guarantee the convergence to the correct solution. These terms $(n\neq0)$ also contribute to the effective spontaneous emission rate, which has been previously calculated only from the $n =0$ term \cite{superduper1, Dinc19, Zheng13}, an issue that requires careful treatment  (see Apeendix\,\ref{App:respeak} for more details).


\subsubsection{Solution in terms of wavepacket multiple reflections}
We can write an alternative solution to the dynamics as follows \cite{Milonni74}
\eqn{\label{Peters}
c_\pm \bkt{t}=\frac{1}{\sqrt{2}}\sum_{n = 0 }^\infty&\sbkt{\frac{\bkt{\mp\beta e^{i \phi_p }}^n}{n!} \bkt{ \frac{\gamma t-n\eta}{2} }^n\right.\non\\
&\left. e^{ - \bkt{ \gamma t - n\eta }/2} \Theta \bkt{ t - \frac{n\eta}{\gamma}}}.
}

The above expansion can be understood in terms of a cascade of processes as the field emitted by each of the atoms propagates back and forth between them at signaling times of $ t = nd/v$. The field then coherently adds to the existing amplitudes, offering the intuition that the decay dynamics arises from the multiple partial reflections of a field wavepacket bouncing between the atoms. 

Although the two solutions in Eq.\,\eqref{fW} and Eq.\,\eqref{Peters} appear different, they are equivalent, providing complementary insights in the dynamics of systems with self-consistent time-delayed feedback~\cite{superduper1}. 


\subsubsection{Effective decay rates in presence of retardation}

From the full solution of the dynamics for the two atoms, we can define an effective instantaneous atomic decay rate for the $m ^\mr{th}$ atom as
\eqn{\label{gammam}
\gamma^\mr{eff}_m \equiv - 2 \re\sbkt{ \frac{1}{c_m(t)} \der{c_m(t)}{t} }.
}

Substituting Eq.\,\eqref{c12} and Eq.\,\eqref{dfdt} in Eq.\,\eqref{gammam}, we find that
\eqn{
\gamma_{1,2}^\mr{eff}\bkt{t} =  \gamma\bkt{1 + \re\sbkt{\beta e^{i \phi_p }\frac{c_{2,1} \bkt{t - \eta/\gamma}}{c_{1,2} (t)}}},
}

The first term corresponds to the individual atomic decay rate, while the second term corresponds to the modification due to a second atom with a delayed interaction, such that its amplitude is evaluated at a retarded time $t\rightarrow t - \eta /\gamma$,  corresponding to the delay time between the two atoms. This illustrates that collective spontaneous emission can be understood as a mutually stimulated emission of two dipoles \cite{Cray82}.

Let us consider the atoms to be initially in a symmetric or anti-symmetric state, with an  amplitude $c_\pm \bkt{t}$ corresponding to the states $\ket{\Psi_\pm }$ ($K_+=0$ or $K_-$=0 respectively). After the field emitted by one atom reaches the other, namely  $d/v<t<2d/v$, the excitation amplitudes for the symmetric and anti-symmetric states can be written from Eq.\,\eqref{c12} and  Eq.\,\eqref{Peters} as
\eqn{\label{csupsub}
&c_\pm \bkt{t}=\frac{1}{\sqrt{2}}\sbkt{e^{- \gamma t/2} \mp\beta e^{i \phi_p } \bkt{ \frac{\gamma t-\eta}{2} }e^{ - \bkt{ \gamma t - \eta }/2} },
}
The corresponding instantaneous rate of spontaneous emission for  $t\rightarrow \frac{d}{v}^+$ is
\eqn{\label{gammaeff}
\lim_{ t \rightarrow \frac{d}{v}^+}\gamma^\mr{eff}_\pm= \gamma \bkt{1\pm \beta  \cos\phi_pe^{\eta /2}}.
}

Further assuming that $\beta \cos \phi_p = 1 $, we note that the instantaneous emission rate for a pair of initially symmetric dipoles  $\gamma_+^\mr{eff}$ can exceed  $2\gamma$, a feature referred to as \textit{superduperradiance} \cite{superduper1}. We also observe from Eq.\,\eqref{gammaeff} that  $\gamma _-^\mr{eff}$ could be negative, illustrating that an initially  asymmetric state can exhibit recoherence, exciting the atoms after they have decayed \footnote{It can be noted that the above instantaneous collective emission rate differs from those quoted in \cite{superduper1, Dinc19, Zheng13} where the contribution from the $n\neq0$ terms in  Eq.\eqref{fW} is neglected.
}.


\subsubsection{Energy shifts from retarded dipole-dipole interaction}

The energy shift due to the retarded resonant dipole-dipole interaction can be defined as \cite{Milonni15}
\eqn{\label{Deltaeff}
\Delta^\mr{eff}_{m}(t)  \equiv -2\im{\sbkt{ c^\ast_\mr{m} (t) \der{c_\mr{m} (t)}{t}}}.
}

For a general initial state the shift at any time $t$ is given as,
\eqn{
\Delta_{1,2}^\mr{eff} \bkt{t} = \gamma\beta \im\sbkt{  e^{i \phi_p} c_{1,2}^\ast\bkt{t}c_{2,1}\bkt{t- \eta/\gamma} }.
}
Considering the two atoms to be initially in a symmetric or anti-symmetric state, the instantaneous shift at $ {\frac{d}{v}}<t<{\frac{2d}{v}}$ can be found from substituting  Eq.\,\eqref{csupsub} in Eq.\,\eqref{Deltaeff}  as
\eqn{\lim_{t\rightarrow {\frac{d}{v}}^+}
\Delta _\pm  ^\mr{eff}(t) = \pm \frac{\gamma}{2}e^{-\eta/2} \beta \sin \phi_p.
\label{eq:shift-decay}
}
Comparing the above with the  effective instantaneous emission rate $\gamma^\mr{eff}_\pm$ in Eq.\,\eqref{gammaeff} we note that the two are related to each other as the real and imaginary part of a common response function \cite{PeterQOBook}. 

We further remark  that  for small distances the dipole-dipole interaction is oscillatory $(\sim \sin \phi_p)$ \cite{Milonni15}, as predicted by a Markov approximation. Upon including  retardation, the dipole-dipole interaction decays exponentially with distance, with the coherence length of the spontaneously emitted photon $(\sim v/\gamma )$ determining the characteristic length scale. This suggests that dipole-dipole interactions in one-dimension cannot be truly infinite-ranged \cite{Solano2017} because of retardation, setting another limit to the range of collective emitter interactions in one-dimensional systems beyond the ones discussed in Ref.~\cite{SanchezBurillo20}.


\subsection{Field dynamics and spectrum}
\label{sec:field}

Having written the atomic dynamics, we now consider the dynamics  of the electromagnetic field modes and the spectrum of the field, which is defined as the probability of finding a photon of frequency $\omega$ in the waveguide modes at any given time. We substitute the solution for atomic dynamics in Eq.\,\eqref{c12} into the equations of motion for the field in Eq.\,\eqref{ca1} and Eq.\,\eqref{cb1} to obtain the field excitation amplitude as follows
\eqn{
c_a\bkt{\omega, t} = - i \sqrt{\frac{\beta\gamma}{\pi}}& \left[ K_+ \cos\bkt{\frac{k d}{2}}F_+(\omega,t)\right.\non\\
&\left.-i K_- \sin\bkt{\frac{k d}{2}}F_-(\omega,t)\right]}

\eqn{ c_b\bkt{\omega, t} = - i \sqrt{\frac{\beta\gamma}{\pi}} &\left[ K_+ \cos\bkt{\frac{k d}{2}}F_+(\omega,t)\right.\non\\
&\left.+i K_- \sin\bkt{\frac{k d}{2}}F_-(\omega,t)\right]}
where we have defined \eqn{\label{Fpm}
F_\pm \bkt{\omega,t} =& \int_0 ^t \dd\tau \,c_\pm\bkt{\tau } e^{i \bkt{\omega - \omega_0 } \tau}. 
}

\begin{figure}[t]
    \centering
    \includegraphics[width = 3.35 in]{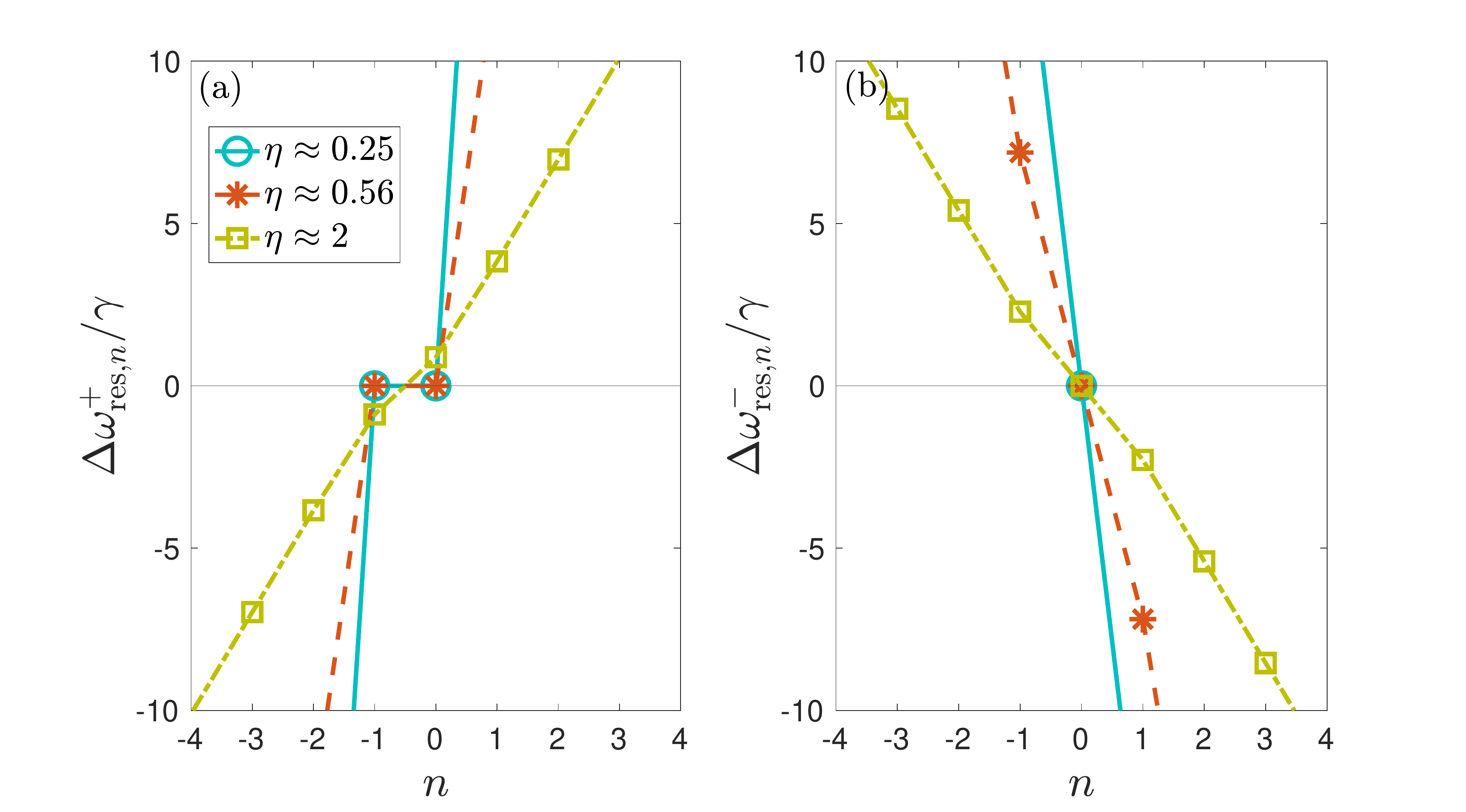}
    \caption{Shift in the resonant frequency $\Delta\omega^\pm_{\mr{res},n}$ corresponding to the different branches of the $W$-function for the initial (a) symmetric and (b) anti-symmetric states. The prominent contributions to the total spectrum are due to $n = \cbkt{-1,0}$ for the symmetric case, and $n = \cbkt{-1,0, 1}$ for the antisymmetric case \cite{superduper1}. It can be seen from (a) that for small enough atomic separations $(\eta <\eta_c)$ there is no shift to the prominent  resonance peaks $(\Delta\omega^+_{\mr{res},n}\approx 0)$  corresponding to $n = \cbkt{-1,0}$. The $\eta $ values are chosen such that $ \phi_p = \eta\omega_0/\gamma = 2p \pi$.}
    \label{omegares}
\end{figure}

\begin{figure*}[t]
\centering
  \subfloat{\includegraphics[width = 3.25 in]{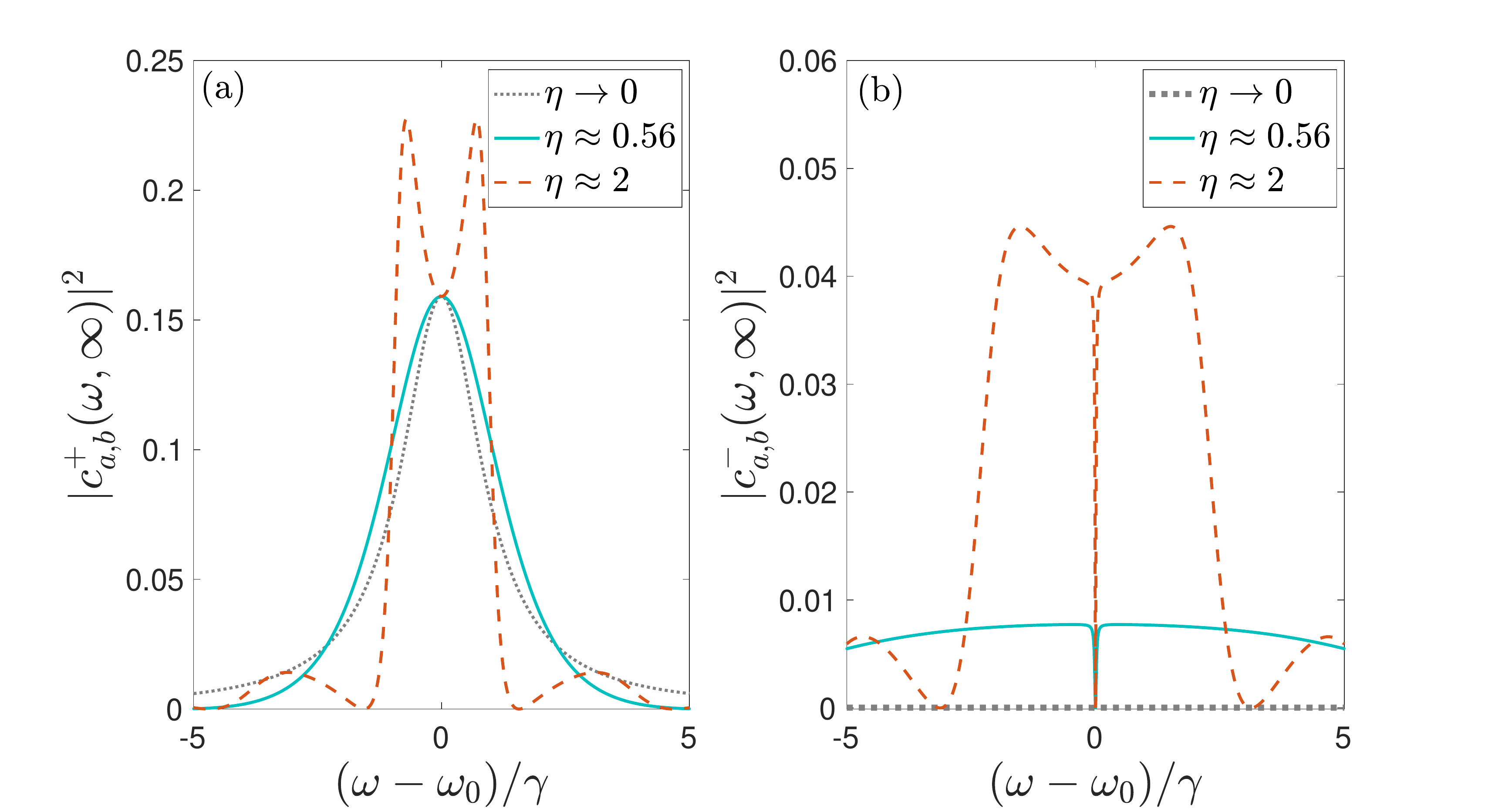}}
  \subfloat{\includegraphics[width = 3.25 in]{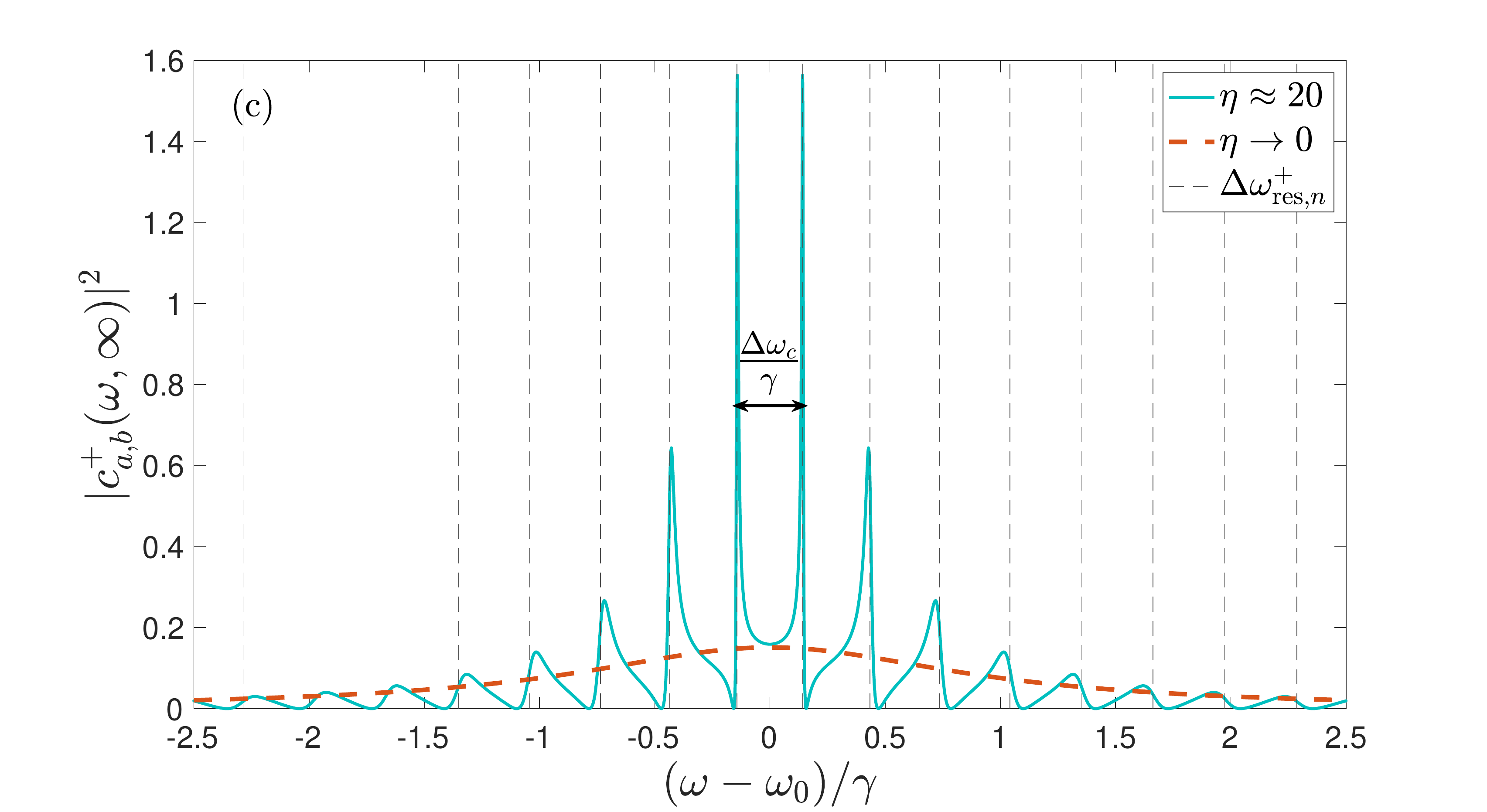}}
    \caption{  Steady state excitation probability of waveguide modes as a function of frequency for different  interatomic separations for initial (a) symmetric and (b) antisymmetric states. It can be deduced numerically from (a) that the FWHM  of the late-time spectrum for $\eta\approx0.56$ is $\Delta_\mr{FWHM}\approx 2.57$ which  exceeds that for  the case of coincident emitters $\eta =0$ ($\Delta_\mr{FWHM}\approx 1.97$, which in the limit of perfect coupling to the waveguide with $\beta\rightarrow1$ approaches $\Delta_\mr{FWHM}\rightarrow2$), this is a signature for retardation-induced modification to collective atomic decay. For $\eta>\eta_c$, there are multiple peaks in the late time spectrum for the initial symmetric state. (c) Late time spectrum for $\eta\approx 20$ for an initial symmetric state is depicted as the solid blue curve. The self phase modulation dynamics in the atomic cavity leads to  Fano resonance-like spectrum. The spacing between each resonant peak is roughly given as $\Delta\omega_c/\gamma  \approx 0.286$ from Eq.\,\eqref{deltawc}. The dashed vertical lines represent the resonance frequency peaks $\Delta\omega_{\mr{res},n}^+$ given in Eq.\,\eqref{wres}. The width of the central resonant peak is given by Eq.\,\eqref{gres} as $\gamma_{\mr{res},0}^+  \approx 0.0083\gamma$.  The red dashed curve corresponds to the superradiant spectrum for two coincident emitters $(\eta \rightarrow0)$. The  $\eta$ values are chosen such that $\phi_p = \eta\omega_0/\gamma = 2 p \pi$. }
    \label{spectrum}
\end{figure*}

In the late time limit, $t\rightarrow\infty$, we obtain
\eqn{\label{FpminfW}F^{\infty}_\pm (\omega) = \frac{i}{\sqrt{2}} \sum _{n\in \mathbb{Z}}\frac{\alpha _n ^{(\pm)} }{ \sbkt{\bkt{\omega -\omega_0} -\Delta \omega_{\mr{res},n}^{\pm} } +i \gamma_{\mr{res},n} ^{\pm}/2},}
which illustrates that there are multiple resonant peaks in the outgoing field spectrum at frequencies $\bkt{\omega_0 + \Delta\omega_{\mr{res}, n}^\pm}$ with a corresponding width $\gamma_{\mr{res}, n}^\pm$. The resonance frequency shifts $\Delta\omega_{\mr{res},n}^\pm$ and widths $\gamma_{\mr{res},n}$ are given by
\eqn{ \label{wres}
\Delta\omega_{\mr{res},n}^\pm &= - \frac{\gamma} {\eta} \im \sbkt{W_n\bkt{ \mp \beta e^{i \phi_p } \frac{\eta}{2} e^{\eta/2}}}\\
\label{gres}
\gamma_{\mr{res},n}^\pm &= \gamma\sbkt{1 - \frac{2}{\eta} \re \sbkt{ W_n\bkt{ \mp \beta e^{i \phi_p } \frac{\eta}{2} e^{\eta/2}}}}.
}
Note that the steady state field spectrum is the Fourier transform of the transient atomic dynamics, thus the  lineshifts and linewidths obtained above are in agreement with Eq.\,\eqref{fW}. Fig.\,\ref{omegares} shows the resonance frequencies for different $n $ values. We note from Fig.\,\ref{omegares}\,(a) that for small enough atomic separations $\eta<\eta_c$, there is no shift in the resonance peaks for the symmetric spectrum for $n = 0,-1$, the dominant orders.

Figure\,\ref{spectrum} shows the spectrum of field radiated outside the system. For a value of $\eta<\eta_c$, the spectrum has a single maximum.   This can be understood from the fact  that for small enough atomic separation the late time spectrum has only one resonance at $\omega \approx \omega_0 $ as seen in Fig.\,\ref{spectrum}\,(a). In this  limit the two atoms  behave as a single entity. For distances $\eta>\eta_c$, we see that there are two prominent peaks corresponding to $\Delta\omega_\mr{res,n }^+$ with $n = -1,0$. This corresponds to the limit where the two atoms make an effective cavity and interact with the field oscillating within. In the case of a symmetric atomic state (see Fig.\,\ref{spectrum}\,(a)), the spectral width of the spectrum as defined via its full width at half-maximum (FWHM) increases as a signature of the enhancement of the spontaneous emission beyond Dicke superradiance.

In the case of an anti-symmetric atomic state (see Fig. \ref{spectrum}\,(b)), the spectrum is ideally zero, since a perfect subradiant states does not radiate light for $\eta =0$. However, for $\eta>0$, a broad spectrum appears from the field that leaks out of the system before the atoms interact with each other, suddenly turning it off once the atoms ``see'' each other and destructively interfere. This turn-off process contributes with a broad range of frequency components. In the case of anti-symmetric emitters, there is always a narrow dip at $\omega=\omega_0$. This is because the resonant radiation into the waveguide is mostly trapped in the region between the emitters, but over time it can be scattered out into external modes. The dip in the center is therefore determined by the waveguide coupling efficiency $\beta$.

In the long cavity limit, meaning $\eta>1$, the field wavepacket radiated by the atoms is reflected multiple times within the effective cavity formed by the atoms. The resulting output field is a train of  pulses separated by a time $d/v$. In the frequency domain this results in multiple resonance peaks as seen in Fig.\,\ref{spectrum}(c).  Each resonant peak corresponds to a Lorentzian in Eq.\,\eqref{FpminfW}, that are all phase coherent with each other. The asymmetry of each peak arises from a Fano-like interference between the atomic resonance and the resonances of the cavity made by the atoms \cite{Fano}.

For the case of a symmetric initial state, the separation between the central resonant peak is given by
\eqn{\label{deltawc}
\Delta\omega_c =  \frac{2\gamma} {\eta}\im\sbkt{W_0 \bkt{- \beta \frac{\eta}{2} e^{\eta/2}}}.
}
As the two atoms are separated infinitely further apart, $\lim _{\eta \rightarrow\infty}\im \sbkt{ W_0 \bkt{ - \beta \frac{\eta}{2}e^{\eta /2}} } = \pi $. Thus the spacing between different teeth asymptotically approaches  ${\Delta \omega_c}/\bkt{2\pi} \rightarrow  v/d$.  This is twice the free spectral range of a  Fabry-Perot cavity of length $d$, with  ${\Delta \omega_\mr{FSR} }/\bkt{2\pi } = v/(2d)$ \cite{PierreBook}. This can be understood from the fact that while for a cavity the field leaks out after every round trip time $( 2d/v)$, in the two atom system, the field leaks out of the atomic ``cavity'' after every half-round trip $ d/v$. The correspondence to a cavity only applies as an asymptotic behaviour when the atoms are placed far apart, but in a general scenario the free spectral range in Eq.\,\eqref{deltawc} is determined by the delayed feedback effects between the two atoms.

The late time spectrum can be alternatively written in a more physically intuitive form by substituting  the series solution for the atomic dynamics given in  Eq.\,\eqref{Peters} into Eq.\,\eqref{Fpm} as (see Appendix\,\ref{latetime} for proof)

\begin{widetext}
\eqn{\label{Fpminf} &F_\pm^{\infty}(\omega)= \frac{i}{\sqrt 2}\frac{1}{  \sbkt{\bkt{\omega - \omega_0}  \mp \frac{\gamma\beta  }{2}\sin \bkt{ \frac{\omega \eta}{\gamma}}}+i \frac{\gamma}{2}\sbkt{ 1\pm \beta \cos \bkt{ \frac{\omega \eta}{\gamma}}}  }.
}
\end{widetext}

In the limit of coincident atoms, $\eta\rightarrow 0$, $F_\pm^{\infty} \bkt{\omega}\sim \frac{1}{\bkt{\omega - \omega_0 }+ i \frac{\gamma}{2} \bkt{1\pm \beta}}$, we recover a Lorentzian spectrum for the field \cite{PierreBook}. This gives us the expected Dicke super- and sub-radiant emission profiles for $\beta\rightarrow1$ with a spectrum peak at $\omega = \omega_0$ and linewidths $\gamma^\pm =  2\gamma, 0$.  Deviations from the Lorentzian profile are yet another signature that the dynamics is non-Markovian due to the retardation effects.  It can be seen that the usual  Lorentzian spectrum of the atoms in the Markovian limit is modified by a frequency dependent phase modulation factor $\sim \frac{\gamma\beta}{2} e^{i\phi_p}$ $(\phi_p = \omega \eta/\gamma)$,  the real part of which $\bkt{\frac{\gamma\beta}{2} \cos \bkt{\omega\eta/\gamma}}$ contributes to the linewidth modification and the imaginary part $\bkt{\frac{\gamma\beta}{2} \sin \bkt{\omega\eta/\gamma}}$ to the resonant lineshift. This can be understood as coming from the propagation phase $\phi_p$ for the field modes as they traverse the interatomic distance and interfere with their time-delayed amplitudes. One can also derive the resonant peaks $\bkt{\omega_0+\Delta \omega^\pm_{\mr{res},n}}$ and corresponding linewidths $\gamma^\pm_{\mr{res},n}$ (as in Eq.\,\eqref{wres} and Eq.\,\eqref{gres}) from Eq.\,\eqref{Fpminf} as shown in Appendix \ref{App:respeak}. The above spectrum is also similar to that emitted from a single excited atom placed in  front of a mirror in a retarded regime \cite{Dorner02}. The two problems correspond to each other via image theory.

\section{Driven dynamics}
\label{drive}

We now add a weak drive to the atomic system as given by the  interaction Hamiltonian $\tilde H_{\text{AD}}$ in  Eq.\,\eqref{had} to address the collective state preparation.  Notice that the drive Hamiltonian does not conserve the total number of excitations in the atom+field system. Since solving  the equations of motion for multiple excitations in the presence of non-Markovian feedback is analytically hard,  we solve the driven dynamics perturbatively within the linear regime \cite{Dorner02}, assuming that the Rabi frequency is sufficiently small $(\Omega \lesssim \gamma)$.

Let us assume that the atoms are initially in the ground state and the field in the waveguide is in vacuum, as
\eqn{
\ket{\Psi_0 } \equiv \ket{gg}\otimes \ket{\cbkt{0}}.
}
Considering that the interaction with the drive is switched on at $t = 0$ we can write the amplitudes of excitation for the symmetric and anti-symmetric states $\ket{\Psi_\pm}$ as $c_\pm^D \bkt{t}$ at a time $t$ from first order perturbation theory, obtaining
\eqn{
c_{\pm}^D\bkt{t} &= - \frac{i}{\hbar } \int _0 ^t\dd\tau \bra{\Psi_{\pm}} e^{i \tilde H_{\text{AF}} \tau/\hbar } \tilde {H}_{\text{AD}} e^{-i \tilde H_{\text{AF}} \tau /\hbar}\ket{\Psi_{0}}\non\\
&= - i\sqrt{2}\Omega \int _0 ^t\dd\tau\,   e^{ i\bkt{ \omega_0 - \omega _D }\tau} \bra{\Psi_{\pm}} e^{i \tilde H_{\text{AF}} \tau/\hbar }\ket{\Psi_+ }.
\label{cpmd}
}
This implies that the drive perturbatively excites the atoms into a single excitation symmetric state $\ket{\Psi_+}$ by the virtue of a weak symmetric coupling (see Eq.\,\eqref{had}),  which then evolves in the linear regime via the atom-field interaction Hamiltonian $\tilde H_{\text{AF}}$\footnote{It can be seen from Eq.\,\eqref{c12} that $\avg{ \Psi _- \abs{ e^{i \tilde H_{\text{AF}} \tau }} \Psi_+ } = 0$, meaning that the probability of exciting the antisymmetric state with a symmetric coupling to the drive is zero.}.  We note however that while the atomic state is symmetric, for a propagation phase $\phi_p= (2p+1) \pi $, the two atoms behave subradiantly  \cite{ChangNJP, Mirhosseini19}. In the presence of retardation effects such subradiant states can evolve into highly delocalized entangled states such as the BIC states \cite{superduper1, SPIE, Calajo19}.

\begin{figure}[t]
    \centering
    \includegraphics[width = 3.35 in]{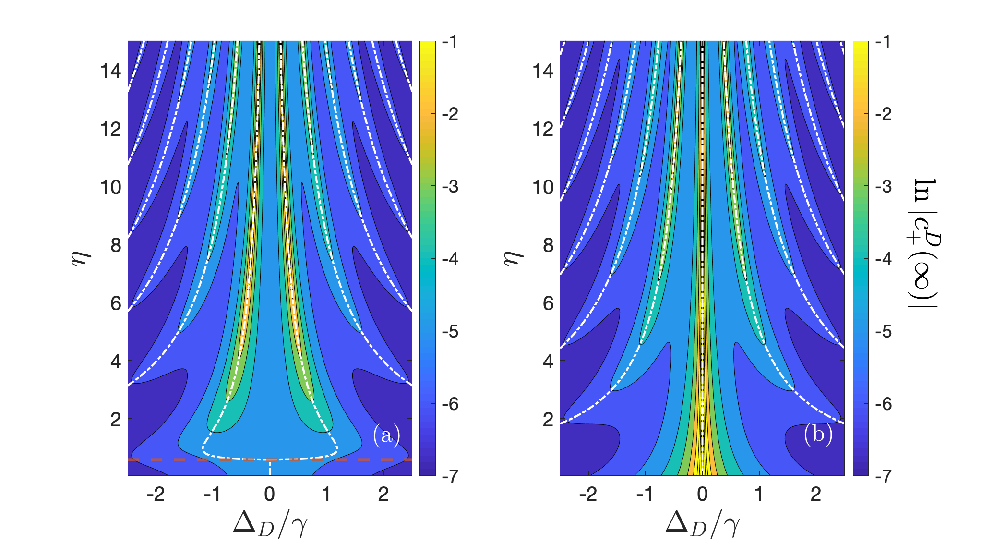}
    \caption{Late time excitation probability for the symmetric state as a function of the drive  detuning and atomic  separation.  The $\eta$ values are chosen such that for (a) $\phi_p = \eta\omega_0/\gamma = 2 p \pi$, which corresponds to the superradiant state and (b) $\phi_p = \eta\omega_0/\gamma= (2 p+1) \pi$ corresponding to the subradiant state. The steady state atomic excitation probability as a function of the drive detuning has multiple peaks at $\omega_D = \omega_0 -\Delta \omega_{\mr{res},n}^+$ as determined by   Eq.\,\eqref{wres}, as depicted by the dashed-dotted white curves. As the atomic separation increases the number of peaks increases similar to the Fano resonance-like structure of the late time spectrum in Fig.\,\ref{spectrum}\,(c). The red dashed line in (a) represnts the critical distance $\eta = \eta_c $ after which there is a bifurcation of the central peak. It can be seen that in a subradiant  configuration Rabi frequency for the weak driving field is  $\Omega= 0.1 \gamma$.}
    \label{Fig:drivenatom}
\end{figure}

The steady state excitation amplitude for the symmetric state can be thus obtained from Eq.\eqref{cpmd}
\eqn{\label{c+Dinfw}c_+^{D}\bkt{t\rightarrow\infty} = \Omega \sum_n  \,\frac{ {\alpha_n ^{(+) }}^\ast}{(\omega_D -\omega_0 + \Delta \omega^+_{\mr{res}, n}) +i \gamma^+_{\mr{res}, n}  },}
 where $\Delta \omega_{\mr{res}, n }^+$ and $\gamma_{\mr{res}, n}^+$ are as defined in  Eq.\,\eqref{wres} and Eq.\,\eqref{gres} respectively. Thus we note that there are multiple peaks in the steady state atomic amplitude for a drive frequency such that $\omega _D = \omega_0 - \Delta  \omega_ {\mr{res},n}^+$, with a corresponding width of $\gamma_{\mr{res},n}^+$. Similar to Eq.\,\eqref{Fpminf}, this  can be alternatively  written as
\eqn{
&c_+ ^D \bkt{t\rightarrow\infty} \non\\
&= \frac{ \Omega}{\sbkt{\Delta_D   + \beta \frac{\gamma}{2} \sin \bkt{ \frac{\omega_D \eta}{\gamma}}}+ i\frac{\gamma}{2}\sbkt{ 1+ \beta \cos \bkt{ \frac{\omega_D \eta}{\gamma}}}  },
}
where $\Delta _D = \omega_0 - \omega_D $ is the drive detuning. In the limit of the two emitters being coincident, we find that the excitation amplitude $c_+ ^D \bkt{t\rightarrow\infty } \rightarrow \frac{ \Omega}{\Delta_D + i\frac{\gamma}{2} \bkt{1+ \beta}  }$ is described by the familiar Lorentzian profile as a function of the detuning $\Delta_D$ with a width  $\gamma \bkt{1+\beta}/2$ \cite{PierreBook}.

Figure \,\ref{Fig:drivenatom}  shows the symmetric state excitation probability in the late time limit as a function of the drive detuning and atomic separation, for the specific propagation phases of $\phi_p = 2p \pi , (2p+1)\pi$ which correspond to a superradiant and a subradiant pair of emitters respectively. We observe multiple peaks in the excitation probability corresponding to drive frequencies $\omega _D = \omega_ 0 - \Delta \omega_{\mr{res},n}^+$ as determined by Eq.\,\eqref{c+Dinfw}. 

This scheme can be used to prepare two distant atoms in an entangled state, depending upon the drive detuning and the atomic separation. We remark however that these results are limited by the applicability of perturbation theory and require the drive strength to be sufficiently weak. The dynamics of the scattered field is discussed in Appendix\,\ref{rfspectrum}. While for weak driving one only sees the elastic scattering process, in the general case of a strong drive, this would correspond to the resonance fluorescence spectrum of two emitters with retarded interaction \cite{Lenz93}.

\section{Superconducting circuit implementation}
\label{superconducting}

The model and results described here can be implemented in a  circuit QED (cQED) setup. Specifically, we consider a setup with two transmon qubits with resonance frequency $\omega_0 \approx 5 $\,GHz coupled to a Josephson junction (JJ) array as shown in Fig.\,\ref{schematic}(b). We further assume that the two qubits are driven simultaneously by an external pump that couples symmetrically to both. We describe the details of the JJ array in Appendix  \ref{JJA}, and summarize the parameters  value in Table\,\ref{parameters}. For such values, as detailed in Appendix\,\ref{JJA}, a distance of $d \approx 1.6$\,cm  between the qubits corresponds to an  $\eta \approx 1$, where the system would exhibit significant retardation effects. These parameters  are within the fabrication capabilities of ongoing experiments \cite{Leger19, Kuzmin19}.

Collective effects in  cQED have been already observed in a system of two artificial atoms coupled to a microwave cavity \cite{vanloo13, Mlynek14}, and recent implementations have extended their study to multi-qubit systems \cite{Mirhosseini19,Wang20}. Moreover, waveguides made of JJ arrays with low dissipation losses can significantly decrease the field velocity~\cite{Masluk12,  Kuzmin19}.  It is within the reach of current experiments to put both elements together and demonstrate collective atom-field dynamics with retardation effects. 

\begin{center}
\begin{table}[H]
\begin{tabular}{|p {6 cm} |p {2 cm}|}
\hline
{Qubit resonance frequency $\omega_0/(2\pi)$}  &  5\,GHz  \\
\hline
{Decay rate $\gamma/(2\pi)$ } & 10\,MHz  \\
\hline
{Waveguide coupling efficiency $\beta$ } & 0.95\\
\hline
{Phase velocity $v/c$ } & 1/300\\
\hline
\end{tabular}
\caption{Parameter values for a superconducting circuit implementation of the model as depicted in Fig.\,\ref{schematic}\,(b).  We use this parameters throughout the paper to present our results under realistic conditions.}
\label{parameters}
\end{table}
\end{center}
\section{Summary and Discussion}
\label{discussion}
To summarize, we study a system of two driven distant emitters coupled to a one-dimensional waveguide considering retardation effects. We analytically solve the dynamics of the system for a general initial atomic state in the single-excitation subspace.  We illustrate the collective atomic decay as mutually-induced stimulated emission process (see Eq.\,\eqref{gammaeff}). We also show that dipole-dipole interactions decay exponentially as a function of the interatomic distance, with a characteristic length scale determined by the linewidth of the field (see Eq.\,\eqref{eq:shift-decay}). We find that the spectrum of the radiated field can exhibit a linewidth broadening beyond that of standard Dicke superradiance (Fig.\,\ref{spectrum}(a)). Additionally, if the atoms are widely separated, the spectrum of the field exhibits Fano resonance-like peaks, shown in Fig.\,\ref{spectrum}(b). We finally consider a weak drive in the system, to prepare entangled atomic steady states, and determine the parameters of the drive that allow the preparation of a particular collective state. We further illustrate that one can realize the model in a cQED implementation, with parameter values within reach of the state-of-the-art setups.

As an outlook, this work represents a step forward towards the study of strongly driven dynamics in the retarded regime. While we have explored the dynamics in a single-excitation regime, its extension to multiple excitations in the system, can exhibit non-linear and distinctly quantum features. As an example of such regime, one can consider the modifications to the resonance fluorescence spectrum of two atoms due to retardation \cite{Lenz93}. Given the present analysis of two emitters coupled to a waveguide, the study of multiple emitters seems a natural extension, as recently explored in Ref. \cite{Dinc19} by studying the atomic dynamics. It would be interesting to explore the linewidth and coherence properties of the narrow frequency  peaks in the radiation spectrum in the many-atom scenario \cite{Khan18}. Additionally, circuit QED setups with tunable qubit frequencies and engineerable JJ arrays allow for implementing a determined spectral density of modes that can help with efficient steady state entanglement generation between the qubits \cite{Hartmann08, Zueco12}.

\section{Acknowledgement}
We are specially grateful to Pierre Meystre for inspiring discussions throughout the course of the work. We also thank Peter W. Milonni,  Luis A. Orozco, Elizabeth A. Goldschmidt, Hakan E. T\"{u}reci,  and Saeed A. Khan for  insightful comments and thoughtful reading of the manuscript.
K.S acknowledges helpful discussions with Ahreum Lee, Hyok Sang Han, Fredrik K. Fatemi and S. L. Rolston. A. G.-T. acknowledges support from the Spanish project PGC2018-094792-B-100 (MCIU/AEI/FEDER, EU) and from the CSIC Research Platform on Quantum Technologies PTI-001. Y.L. acknowledges the Knut, Alice Wallenberg Foundation, and the Swedish Research Council for financial support. P.S. was supported by CONICYT-PAI grants 77190033. K.S. acknowledges support from the US Department of Energy, Office of Basic Energy Sciences, Division of Materials Sciences and Engineering, under Award No. DE-SC0016011.

\appendix

\section{Late time field dynamics}

\label{latetime}
Substituting  the series solution for the atomic dynamics  Eq.\,\eqref{Peters} in Eq.\,\eqref{Fpm}
\begin{widetext}
\eqn{F_\pm \bkt{t\rightarrow\infty} =&\lim_{t\rightarrow\infty}\frac{1}{\sqrt{2}} \int _0 ^t \dd\tau e^{i \bkt{\omega - \omega_0 }\tau } \sum_{n = 0 }^\infty\frac{\bkt{\mp\beta e^{i \phi_p }}^n}{n!} \bkt{ \frac{\gamma \tau-n\eta}{2} }^ne^{ - \bkt{ \gamma \tau - n\eta }/2} \Theta \bkt{ \gamma \tau - n\eta}.}
Let us define $\tilde \tau  \equiv \frac{\gamma \tau - n \eta }{2}$, to rewrite the above as
\eqn{F_\pm \bkt{t\rightarrow\infty} = \frac{1}{\sqrt{2}} \sum_{n = 0 }^\infty \frac{\bkt{\mp\beta e^{i \phi_p }}^n}{n!} \bkt{\frac{2}{\gamma}e^{i \bkt{\omega - \omega_0 }n \eta/\gamma}}\sbkt{\int _0 ^\infty \dd\tilde \tau \tilde  \tau^n e^{- \tilde \tau} \Theta \bkt{\tilde \tau}e^{2i \bkt{\omega - \omega_0 }\tilde \tau /\gamma } }.
}

\end{widetext}
Now using the Laplace transform identity $\int _0 ^\infty\dd x e^{- s x } x ^n e^{- \alpha x} \Theta \bkt{x } = \frac{n !}{(s + \alpha )^{n+1}} ,$ we can simplify the integral in the square bracket above to obtain

\eqn{
&F_\pm \bkt{t\rightarrow\infty} =\non\\
& \frac{1}{\sqrt{2} \sbkt{ \frac{\gamma}{2}-i \bkt{\omega - \omega_0 }}} \sum_{n = 0 }^\infty \sbkt{\mp\frac{\beta e^{i{\omega \eta/\gamma} } }{\cbkt{1-2i \bkt{\omega - \omega_0 }/\gamma }}}^n
}
which yields Eq.\,\eqref{Fpminf} using the identity $\sum_{n = 0 }^\infty x^n = \frac{1}{1 - x}$, for $\abs{x}<1$, which is ensured from the coupling efficiency $\beta$ being less than 1.

\section{Resonant peaks in late time spectrum}
\label{App:respeak}

Let us consider the characteristic equation for resonant peaks from the denominator in Eq.\,\eqref{Fpminf}
\eqn{\label{chareq}
\bar\omega^\pm - \omega_0 + i \frac{\gamma}{2} \pm i \frac{\gamma\beta}{2}e^{i \bar\omega \eta /\gamma}=0.
}
Defining  $\tilde \omega^\pm = \bkt{\bar\omega^\pm - \omega_0 +i \frac{\gamma}{2}}$,
\eqn{
&\tilde  \omega^\pm \pm i \frac{ \gamma\beta}{2} e^{i \eta/\gamma\sbkt{\tilde  \omega  + \omega _0 - i \gamma/2}} =0\\
\implies& \bkt{- i  \frac{\tilde\omega\eta }{\gamma}}  e^{-i \tilde \omega\eta/\gamma}= \mp \beta e^{i \phi_p }\frac{ \eta}{2} e^{\eta /2}\\
\implies& -i \frac{\tilde \omega_n \eta }{\gamma} = W_n\bkt{\mp \beta e^{i \phi_p }\frac{ \eta}{2} e^{\eta /2}},
}
where we have used the definition of the Lambert-$W$ function that $W (x) = f^{-1}(x) $, with  $f(x) = xe^x$. This yields the complex eigenvalue of the characteristic equation as
\eqn{\bar \omega_n^\pm=& \omega_0 - i \frac{\gamma}{2}+ i \frac{\gamma}{\eta} W_n\bkt{\mp \beta e^{i \phi_p }\frac{ \eta}{2} e^{\eta /2}}\\
=& \omega_0 -i \frac{\gamma_n ^{(\pm)}}{2},
}
where we have used Eq.\,\eqref{gammanpm}. The real and imaginary part of the above yields the resonant peak frequencies $\omega_0 + \Delta\omega_{\mr{res},n}^\pm$ and the corresponding linewidths $\gamma_{\mr{res},n}^\pm$ of the late time spectrum as in Eq.\,\eqref{wres} and \eqref{gres}.

We further remark that Eq.\,\eqref{chareq} is the characteristic equation used in solving the atomic dynamics in Refs. \cite{Dinc19, Zheng13}, though in these works only the zeroth eigenvalue is considered. This gives an effective decay rate that corresponds to only $\gamma^\pm_{\mr{res}, n=0}$. However the full solution to the dynamics includes other eigenvalues as well, and yields an effective decay rate that differs from \cite{superduper1, Dinc19, Zheng13}.

\section{Josephson junction array as waveguide}
\label{JJA}
\begin{figure}[t]
    \centering
    \includegraphics[width =2.5 in]{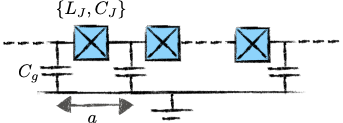}
    \caption{Schematic representation of a JJ array circuit. Each JJ, represented in blue is considered to be a linear $LC$-oscillator with an inductance  $L_J\approx1$\,nH and capacitance $C_J\approx 1$\,fF, and is connected to the ground with a capacitance $C_g\approx100$\,fF. Each unit cell is of length $a\approx 10 $\,$\mu$m.}
    \label{JJAfig}
\end{figure}

We consider two transmon qubits with a resonance frequency $\omega_0$ coupled to a JJ array made of $N \approx 2000$ identical JJs in series, as shown in Fig.\,\ref{schematic}\,(b). 
The schematic  for a  JJ array is depicted in  Fig.\,\ref{JJAfig}. Assuming that the JJs are linear such that each JJ can be treated as an LC oscillator, one can derive the dispersion relation for such a waveguide, following the approach in \cite{Masluk12} such that

\begin{figure}[b]
    \centering
    \includegraphics[width = 3.35 in]{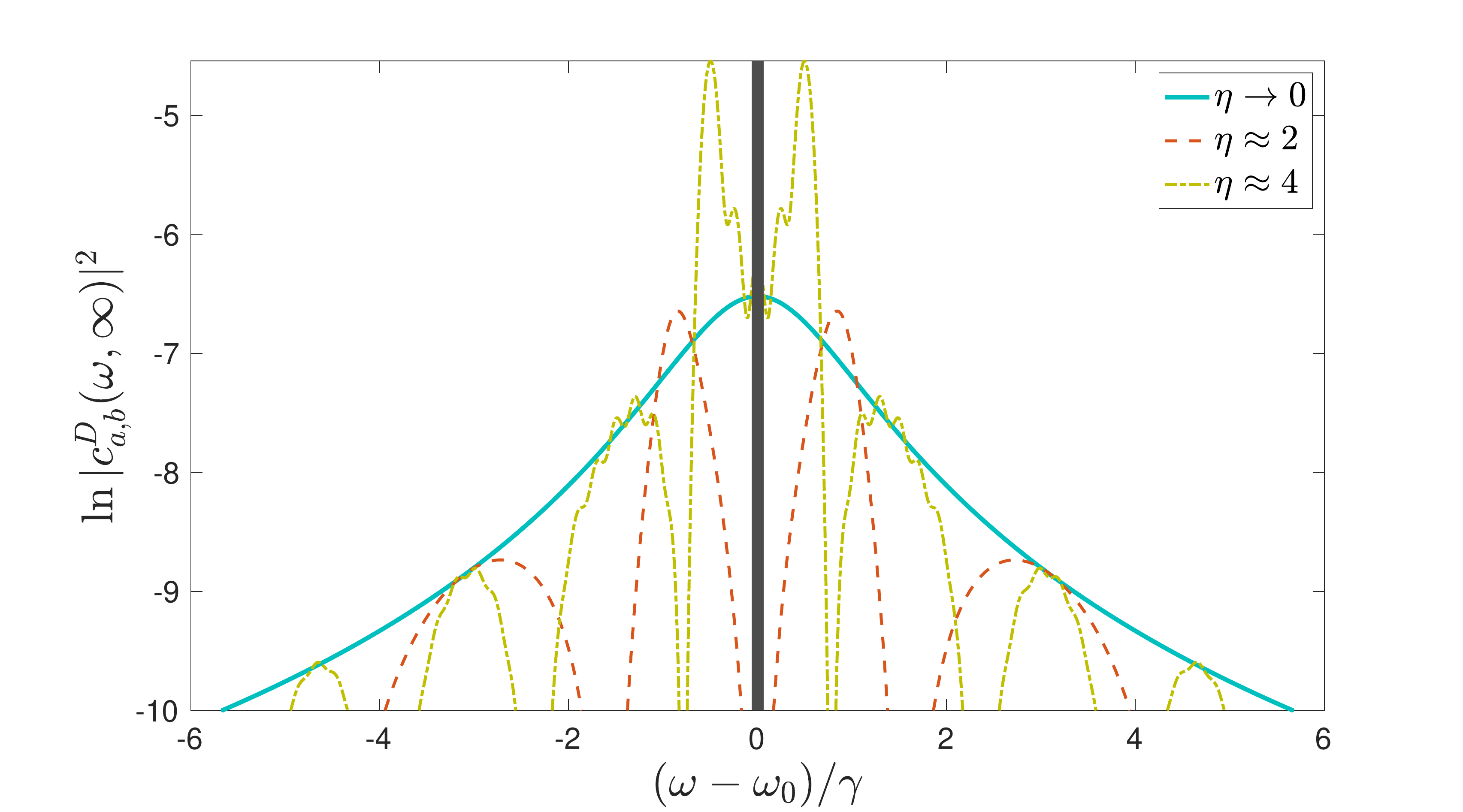}
    \caption{Scattering spectrum from a pair of driven emitters at different separations. The propagation phase is assumed to be $\phi_p = 2 p \pi$. The solid line at $\omega = \omega_0$ denotes that the resonant frequency is being filtered.}
    \label{Figrf}
\end{figure}
\eqn{
\omega(k) = \frac{1}{\sqrt{L_JC_J}} \sqrt{\frac{1 - \cos\bkt{ka}}{\frac{C_g}{2C_J}+ \bkt{1 - \cos\bkt{ka}}}},
}
where $ka = n\pi  /N$. With the chosen parameter values for $L_J$, $C_J$ and $C_g$ as indicated in Fig.\,\ref{JJAfig}, the phase velocity of the field around  $\omega_0 \approx 5 $\,GHz is $v = \omega/k \approx 1\times 10^6$\,m/s.

\section{Driven field spectrum with retardation}
\label{rfspectrum}
The driven symmetric state amplitude can be obtained from Eq.\,\eqref{cpmd} as
\eqn{
\label{c+dt}
c_+  ^D\bkt{t}  =& -  \Omega\sum_{n\in\mathbb{Z}} {\alpha_n ^{(+)}}^\ast\frac{ e^{\sbkt{i\Delta_D -{\gamma_n ^{(+)}}^\ast/2}t} - 1}{\Delta_D +i{\gamma_n ^{(+)}}^\ast/2},
}
where  $\Delta_D \equiv \omega_ 0 - \omega_D$ is the  detuning of the laser with respect to the atomic resonance. We note that the above expression for the atomic excitation amplitude is similar to that for the field amplitude in Eq.\,\eqref{Fpm} with  $\omega  \rightarrow\omega_D$, and thus exhibits the similar features as those in Section\,\ref{sec:field}.

We consider  the dynamics of the waveguide  modes as sourced by the two atoms. The excitation amplitudes of the field modes are obtained by substituting Eq.\,\eqref{c+dt} in Eq.\,\eqref{ca1} and Eq.\,\eqref{cb1}, and subsequently integrating those to yield


\eqn{
c_{a,b}^D\bkt{\omega , t} =& - \Omega \sqrt{\frac{\beta \gamma }{\pi}} \cos \bkt{\frac{k d }{2}} \sum_n \frac{{\alpha _n ^{(+)}}^\ast}{ i \Delta_D - {\gamma_n ^{(+)}}^\ast/2}\non\\
&\sbkt{ \frac{e^{\cbkt{i \bkt{\omega - \omega_D}   - {\gamma_n ^{(+)}}^\ast/2} t}-1}{i \bkt{\omega - \omega_D}   - {\gamma_n ^{(+)}}^\ast/2 } - \frac{e^{{i \bkt{\omega - \omega_0}   } t}-1}{i \bkt{\omega - \omega_0}    }}
}

The first term in the above expression is due to the field emitted from symmetric state transient dynamics, while the second term corresponds to the field emitted in the steady state. Given the weak driving assumption, this corresponds to only the elastic scattering process in the resonance fluorescence spectrum \cite{Lenz93}.

We consider the two terms in the above separately in the steady state. It can be shown that in the late time limit, the second term corresponds to an infinitely sharp resonant peak, as $\lim_{t\rightarrow\infty} \frac{e^{i \bkt{\omega - \omega_0 }t}-1}{\omega-\omega_0 }\sim \delta \bkt{\omega- \omega_0}$. Assuming that the resonant peak can be filtered, the first term in the steady state limit becomes

\eqn{
c_{a,b}^D\bkt{\omega , t}&= -\Omega \sqrt{\frac{\beta \gamma }{\pi}} \cos \bkt{\frac{k d }{2}} \non\\
& \sum_n \sbkt{\frac{1}{ \Delta_D +i {\gamma_n ^{(+)}}^\ast/2}}\frac{{\alpha _n ^{(+)}}^\ast}{ { \bkt{\omega - \omega_D}   +i {\gamma_n ^{(+)}}^\ast/2 } }.
}
Thus we note from the above that the scattered field has resonant peaks at $\omega = \omega_D + \Delta \omega_{\mr{res},n}^+$, with corresponding widths $\gamma_{\mr{res},n}^+$.  The scattered field spectrum is plotted in Fig.\,\ref{Figrf}. The multiple peaks for a pair of distant emitters are a signature of 
retardation effects.

\end{document}